\documentclass[a4paper,11pt]{article}
\usepackage[utf8]{inputenc}
\usepackage{amsfonts}
\usepackage{amssymb}
\usepackage{graphicx}
\usepackage{fancyhdr}
\usepackage{fancyvrb}
\usepackage{fancybox}
\usepackage{slashbox}
\usepackage{float}
\usepackage{geometry}
\usepackage{minitoc}
 \usepackage{indentfirst}
 \usepackage{multicol}
 \usepackage[outerbars]{changebar}
 \usepackage{pifont,textcomp}
 \usepackage{amsfonts}
 \usepackage{amsmath}
 \usepackage{amscd}
 \usepackage{array}
\usepackage{multirow}
 \usepackage{mathrsfs}
 \usepackage{amssymb}
 \usepackage{amsthm}

\title{ Path integral  in position-deformed Heisenberg algebra with strong quantum gravitational measurement
	   }
\author{ Latévi M. Lawson$^{1,2(a)}$, Prince K. Osei$^{1,3(b)}$, Komi Sodoga$^{2(c)}$ 
	and Fred Soglohu$^{1,3(d)}$
\space\\\\
 ${}^1$ African Institute for Mathematical Sciences (AIMS) Ghana\\
Summerhill Estates, East Legon Hills, Santoe, Accra\\
P.O. Box LG DTD 20046, Legon, Accra, Ghana\\\\
  ${}^2$Université de Lomé, Faculté des Sciences, Departement de Physique,\\
Laboratoire de Physique des Matériaux et des Composants\\
à Semi-Conducteurs, 01 BP 1515 Lomé, Togo\\\\
${}^3$Department of Mathematics
P.O. Box LG 62 
University of Ghana, Legon,Ghana\\\\
latevi@aims.edu.gh$^{a}$, posei@nexteinstein.org$^b$, antoinekomisodoga@gmail.com$^{c}$\\ and  fred@aims.edu.gh$^d$
}

\begin{document}
\maketitle

\begin{abstract}
	Position-deformed Heisenberg algebra with maximal length uncertainty has recently been proven to induce strong quantum gravitational fields at the Planck scale  (2022 J. Phys. A: Math. Theor. 55 105303). In the present study, we use the position space representation on the one hand and the Fourier transform and its inverse representations on the other to construct propagators of path integrals within this deformed algebra. The propagators and the corresponding actions of a free particle and a simple  harmonic oscillator are discussed as examples.	
Since the effects of quantum gravity are strong in this  Euclidean space, we show that the  actions which describe the classical trajectories  of both systems are bounded by the ordinary ones of classical mechanics. This indicates that quantum gravity bends the paths of particles, allowing  them to travel quickly from one point to another. It is numerically observed by the decrease in values of classical actions as one increases the quantum gravitational effects.


\end{abstract}

\section{Introduction}

Due to a constant search for a consistent theory of quantum gravity, the deformation of Heisenberg algebra with minimal measurable uncertainty length has been one of the most intense fields in the last two decades  \cite{3,4,5,6,7,8,9,10,11,12,13,14}. It is widely known that the existence of this minimal uncertainty length presents the issue of high energy requirements that are beyond the scope of any experimental feasibility. To circumvent this requirement, one of us recently has proposed a position-deformed Heisenberg algebra \cite{15} in two dimensions (2D) that introduces the simultaneous existence of minimal and maximal length uncertainties. The emergence of this maximal length demonstrated strong quantum gravitational effects in this space  and predicted the detection of low-energy gravity particles \cite{1}. In continuation of this work, we construct the position space representation describing this maximal length, as well as the corresponding Fourier transform and its inverse representations. We derive the propagators of path integrals based on these different representations by following the work done in minimal length scenarios \cite{16,17,18,19}.

The Hamiltonian's principle of least action is used to generate the  equations of motion. We compute the propagators and actions of a free particle and a simple harmonic oscillator as applications. Since the quantum gravity is strongly measured in this  space \cite{1}, we show that the  classical trajectories of particles described by their actions are deformed, allowing particles to take the shortest path between two points in the minimum time. 
 This result strengthens the claim that the recently proposed position-deformed algebra  \cite{1}  induces strong quantum gravitational fields with features close to those of classical ones of  general relativity.
  
This paper is outlined as follows: in section 2, we establish Hilbert space representations of wave functions associated with this deformed algebra. In section 3, we construct the path integrals in these wave function representations and deduce the corresponding quantum propagators and classical actions. As  examples, we compute the propagators and the actions for some simple models such as the free particle and the harmonic oscillators. In the last section, we present our conclusion.

\section{Position deformed Heisenberg algebra with maximal length} \label{sec3}

Let  $\mathcal{H}=\mathcal{L}^2(\mathbb{R})$ be the
Hilbert space of square integrable functions. The Hermitian  operators $\hat x$ and  $\hat p$ that act in this space     satisfy the condition
\begin{eqnarray}\label{alg1}
	{[\hat x,\hat p ]}=i\hbar\mathbb{I}.
\end{eqnarray}
The corresponding Heisenberg uncertainty principle is given by
\begin{eqnarray}
\Delta x\Delta p\geq \frac{\hbar}{2}.
\end{eqnarray}
Let $ \{|x\rangle\}\in \mathcal{H} $ be   the complete position basis vectors. The action operators in equation (\ref{alg1}) on this basis vector reads as follows
\begin{eqnarray}
	\hat x|x\rangle =x|x\rangle \quad \mbox{and}\quad  \hat p|x\rangle=-i\hbar\frac{d}{dx}|x\rangle,\quad x\in\mathbb{R}.
\end{eqnarray}
The
completeness and orthogonality relations  are given by 
\begin{eqnarray}
	\langle x'|x \rangle =\delta(x-x'),\quad \int_{-\infty}^{+\infty}dx |x\rangle \langle x|= \mathbb{I}.
\end{eqnarray}
Another useful choice of basis vectors is the momentum vector $ \{|p\rangle\}\in \mathcal{H}$ defined by taking Fourier transforms
\begin{eqnarray} \label{Fourier}
	|p \rangle=\int_{-\infty}^{+\infty}dx e^{ipx}|x\rangle \quad \mbox{with}\quad p\in \mathbb{R}
\end{eqnarray}
and  its inverse is defined  as follows 
\begin{eqnarray}
	|x \rangle=\int_{-\infty}^{+\infty}\frac{dp}{2\pi}e^{-ipx}|p\rangle.
\end{eqnarray}
The inner product and completeness relations are given by
\begin{eqnarray}
	\langle p'|p \rangle =\delta(p-p'),\quad \int_{-\infty}^{+\infty}\frac{dp}{2\pi} |p\rangle \langle p|= \mathbb{I}.
\end{eqnarray}
The  action of the operators in (\ref{alg1}) on the vector $ |p\rangle$ is given by
\begin{eqnarray}\label{momentum}
	\hat p |p \rangle=\int_{-\infty}^{+\infty}dx \left(-i\frac{d}{dx} e^{ipx}\right)|x\rangle=	 p |p \rangle,\quad 
	\hat x |p \rangle= \int_{-\infty}^{+\infty}dx \left(i\hbar\frac{d}{dp} e^{ipx}\right)|x\rangle=i\hbar\frac{d}{dp}	|p \rangle.
\end{eqnarray}

%

Let us consider new operators $\hat X$ and $\hat P$  on $\mathcal{H}$. We  define them  by
\begin{eqnarray}\label{op1}
	\hat X=\hat x,\quad 
	\hat P=(\mathbb{I}-\tau \hat x +\tau^2 \hat x^2)\hat p.
\end{eqnarray}
They satisfy the following relation \cite{1}
\begin{eqnarray}\label{alg2}
	{[\hat X,\hat P ]}=i\hbar (\mathbb{I}-\tau \hat X +\tau^2 \hat X^2),
\end{eqnarray}
where  $\tau \in  (0,1)$   is the generalized uncertainty principle parameter related
to quantum gravitational effects in  this space which  describes the Planck scale \cite{3,4}. Obviously by taking $\tau\rightarrow 0$, we recover the algebra (\ref{alg1}). The  action  of these operators on the following unit basis  vectors $ \{|x\rangle\}$ and $ \{|p\rangle\}$  reads  as  follows
 \begin{eqnarray}\label{eq1}
	\hat X |x\rangle&=&x |x\rangle \quad \mbox{and}\quad  \hat P |x\rangle=-i\hbar(1-\tau  x +\tau^2  x^2)\partial_x  |x\rangle, \quad x\in \mathbb{R}.\label{3}\\
	\hat X |p\rangle&=&i\hbar  \partial_p |x\rangle \quad \mbox{and}\quad  \hat P |p\rangle=(1-i\tau\hbar \partial_p  -\tau^2 \hbar^2\partial_p^2 )p  |p\rangle, \quad p\in \mathbb{R}.\label{3}
\end{eqnarray}
Let us  consider an arbitrary vector $|\phi\rangle\in \mathcal{H} $, the projection of this vector on  the unit vectors $ \{|x\rangle\}$ and $ \{|p\rangle\}$    generates the functions $ \phi(x)$ and $ \phi(p)$. As a result, we can write the above equations as follows
\begin{eqnarray}
	\hat X \phi(x)&=&x \phi(x) \quad \mbox{and}\quad  \hat P \phi(x)=-i\hbar(1-\tau  x +\tau^2  x^2)\partial_x\phi(x), \label{3}\\
	\hat X \phi(p)&=&i\hbar  \partial_p \phi(p) \quad \mbox{and}\quad  \hat P\phi(p) =(1-i\tau\hbar \partial_p  -\tau^2 \hbar^2\partial_p^2 )p  \phi(p).
\end{eqnarray}

 An interesting feature  can be observed from  the commutation relation  (\ref{alg2})  through the  following  uncertainty relation:
\begin{eqnarray}\label{uncertitude}
	\Delta  X\Delta P\geq \frac{\hbar}{2}\left(1-\tau\langle \hat X\rangle+\tau^2\langle \hat X^2\rangle\right).\label{in2}
\end{eqnarray}
Using the relation 
$ \langle \hat X^2\rangle=(\Delta  X)^2+\langle \hat X\rangle^2$, the equation (\ref{uncertitude}) can be rewritten as a second order equation for $\Delta X$
	\begin{eqnarray}
	\Delta X^2-\frac{2}{\hbar \tau^2}\Delta P\Delta X +\langle \hat X\rangle^2 -\frac{1}{\tau}\langle \hat X\rangle  +\frac{1}{\tau^2}\leq 0.
\end{eqnarray}	
The solutions for $\Delta X$ are given by 
\begin{equation} \label{equ}
	\Delta X=\frac{\Delta P}{\hbar \tau^2}\pm \sqrt{\left(\frac{\Delta P}{\hbar \tau^2}\right)^2
		-\frac{\langle \hat X\rangle}{\tau}\left(\tau\langle \hat X\rangle-1\right)
		-\frac{1}{\tau^2}}.
\end{equation}
This equation leads to the absolute minimal uncertainty $\Delta P_{min}$ in $P$-direction  and the absolute maximal uncertainty  $\Delta X_{max}$ in $X$-direction  
when $\langle  \hat X\rangle=0$ \cite{15}
\begin{eqnarray}
	\quad \Delta X_{max}=\frac{1}{\tau}\quad  \mbox{and}\quad
	\Delta P_{min}=\hbar\tau.
\end{eqnarray}
It is well known that \cite{3}, the   existence of minimal uncertainty   raises
the question of the loss  of representation  i.e., the space is inevitably bounded by minimal quantity beyond which any further localization of particles is not possible.  In the  presente situation, the minimal  momentum $\Delta P_{min}$   leads to a loss of $\phi(p)$-representation and a  maximal $\phi(x)$-representation. Thus, the corresponding  representation of operators are given by 
\begin{eqnarray} \label{eq19}
	\hat X \phi(x)&=&x \phi(x) \quad \mbox{and}\quad  \hat P \phi(x)=-i\hbar D_x\phi(x) \label{4},
\end{eqnarray}
where $ D_x=(1-\tau  x +\tau^2  x^2)\partial_x $ is a deformed derivative. Using this equation (\ref{eq19}), one can  recover the algebra (\ref{alg2}).

As one can see from the representation of operators in equation (\ref{op1}) or  in equation (\ref{4}), 
the position operator $\hat X$ is Hermitian while the momentum operator $\hat P$ is not
\begin{eqnarray}
	\hat X^\dag=\hat X \quad \mbox{and}\quad 	\hat P^\dag= \hat P+i\hbar\tau(\mathbb{I}-2\tau \hat X) \implies \hat P^\dag\neq \hat P.
\end{eqnarray}
Thus, the Hermicity requirement of the
momentum operator leads to the introduction of the following 
completeness relation \cite{1}
\begin{eqnarray}\label{comp}
	\int_{-\infty}^{+\infty}\frac{dx}{1-\tau  x +\tau^2  x^2}|x\rangle \langle x|=\mathbb{I}. \label{identity}
\end{eqnarray}
To proof the Hermicity of this operator, we have to restrict the  action of $\hat P$ in a physical dense subset,  $\mathcal{D} (\hat P)\subset \mathcal{H}$, which we
shall call the domain of $\hat P$ defined as follows
\begin{eqnarray}
	\mathcal{D}(\hat P)= \{\varphi,-i\hbar D_x\varphi\in \mathcal{L}^2(-\infty,+\infty),\,\,\lim_{x\rightarrow \pm \infty}\varphi(x)=0 \}.
\end{eqnarray}
The restriction to dense subset guarantees the existence of the adjoint
operator $\hat P^\dag$, a necessary condition for one to obtain the Hermicity of this 
operator. The adjoint domain is defined by
\begin{eqnarray}
	\mathcal{D}(\hat P^\dag)= \{\xi,-i\hbar D_x\xi\in \mathcal{L}^2(-\infty,+\infty)\}.
\end{eqnarray}
 Thus, we may write $ \mathcal{D}(\hat P)\subset \mathcal{D}(\hat P^\dag)$, which means that the domain of $\hat P$ is a
proper subset of the domain of its adjoint $\hat P^\dag$. To show the Hermicity of the oparator $\hat P$,  we consider the functional $B(\xi,\varphi)$ defined by 
\begin{eqnarray}
	B(\xi,\varphi):=\langle \xi|\hat P\varphi\rangle -\langle \hat P^\dag \xi| \varphi\rangle.
\end{eqnarray}
Using the relation (\ref{comp}) and  by a straightforward computation of this functional, we have
\begin{eqnarray}
	B(\xi,\varphi)&=& \int_{-\infty}^{+\infty}\frac{dx}{1-\tau  x +\tau^2  x^2}\left[\xi^*(x)\left(-i\hbar D_x\varphi(x)\right)-\left(-i\hbar D_x \xi(x) \right)^*\varphi(x)\right]\cr
	&=& -i\hbar \int_{-\infty}^{+\infty}d\left(\xi^*(x)\varphi(x)\right)=-i\hbar\left[\xi^*(x)\varphi(x)\right]_{-\infty}^{+\infty}.
\end{eqnarray}
Since $\lim_{x\rightarrow\pm\infty}\varphi(x)=0$, and $\xi(x)$ can reach any arbitrary value at the boundaries. This lead to the vanishing of $B(\xi,\varphi)$ i.e., $B(\xi,\varphi)=0$. Consequently, the operator $\hat P$ is a Hermitian  in $\mathcal{D}(\hat P)$ such that
\begin{eqnarray}\label{sym}
\langle \xi|\hat P\varphi\rangle=	\langle \hat P^\dag\xi|\varphi\rangle \implies \hat P=\hat P^\dag.
\end{eqnarray}
 Despite the fact that the momentum is Hermitian, it is not always a self-adjoint operator because its domain includes the domain of $\hat P^\dag$.
 It could have none, or it could have an infinite number of self-adjoint extensions. Note that, as rule in quantum mechanics, the operators that act on square integrable functions are essentially self-adjoint. There are exceptions to the rule. This is because the basic quantization requirement that operators whose expectation values are
 real do not strictly require these operators be self-adjoint. Indeed, the Hermicity result (\ref{sym}) is
 a sufficient condition to ensure that all expectation values of the momentum operator are real. 
Moreover, using the completeness relation (\ref{comp}), the scalar product between two states $|\Psi\rangle$ and $|\Phi\rangle$ and the orthogonality of
 eigenstates become
 \begin{eqnarray}\label{id2}
 	\langle \Psi |\Phi\rangle&=& \int_{-\infty}^{+\infty}\frac{dx}{1-\tau  x +\tau^2  x^2} \Psi^*(x)\Phi(x),\\
 	\langle x |x'\rangle&=&(1-\tau  x +\tau^2  x^2)\delta(x-x') \label{orth1}.
 \end{eqnarray}
 
 To construct a Hilbert
 space representation that describes the maximal length  and the minimal momentum uncertainties, one has to solve the eigenvalue problem
 \begin{eqnarray}
 	-i\hbar D_x \phi_\rho(x)= \rho \phi_\rho (x), \quad\quad \rho\in\mathbb{R}^*.\label{diff}
 \end{eqnarray} 
The solution of this  equation is given by 
 \begin{eqnarray}\label{fzeta}
 	\phi_\rho (x)= A\exp\left(i\frac{2\rho}{\tau\hbar \sqrt{3}}\left[\arctan\left(\frac{2\tau x-1}{\sqrt{3}}\right)
 	+\frac{\pi}{6}\right]\right),
 \end{eqnarray}
 where $A$ is an abritrary constant.
 Then by normalization, $\langle \phi_\rho|\phi_\rho\rangle=1$, we have 
 \begin{eqnarray}\label{nzeta}
 	A&=& \sqrt{\frac{\tau\sqrt{3}}{2\pi}}.
 \end{eqnarray}
 Substituting this equation (\ref{nzeta}) into the equation (\ref{fzeta}) gives 
 \begin{eqnarray} \label{nor}
 	\phi_\rho (x) &=& \sqrt{\frac{\tau\sqrt{3}}{2\pi}} \exp\left(i\frac{2\rho}{\tau\hbar \sqrt{3}}\left[\arctan\left(\frac{2\tau x-1}{\sqrt{3}}\right)
 	+\frac{\pi}{6}\right]\right).
 \end{eqnarray}
 This  wave function describes simultaneously the maximal length  and the minimal momentum uncertainties.
 Furthermore, using the relations (\ref{nor}) and (\ref{orth1}),  we can defined a new  identity operator   as  follows
\begin{eqnarray}\label{id}
 \int_{-\infty }^{+\infty} \frac{d\rho}{\tau\hbar \sqrt{3}}d\rho  |\rho\rangle\langle \rho|=\mathbb{I}.
\end{eqnarray}
This identity operator (\ref{id}) will play the role of the completeness relation of the momentum eigenstates in the derivation of the path-integral.

 By projecting  an arbitrary state $|\psi  \rangle $ onto this  localized states $ |\phi_\rho  \rangle $ one can obtain  the quasi-momentum representation, that is 
\begin{eqnarray}\label{Fourier}
	\psi (\rho)= \langle \phi_\rho|\psi  \rangle 
	=\sqrt{\frac{\tau\sqrt{3}}{2\pi}} \int_{-\infty}^{+\infty}\frac{dx\psi(x)}{1-\tau  x+\tau^2 x^2} 
	e^{-i\frac{2\rho}{\tau \hbar \sqrt{3}}\left[\arctan\left(\frac{2\tau x-1}{\sqrt{3}}\right)
		+\frac{\pi}{6}\right]}.\label{moment}
\end{eqnarray}
This mapping   defined the generalized  Fourier transform of the representation in equation (\ref{nor}).
Its inverse representation is given by 
\begin{eqnarray}\label{Finverse}
	\psi(x)=\frac{1}{\hbar\sqrt{2\pi\tau\sqrt{3}}} \int_{-\infty}^{+\infty}d\rho \psi(\rho)  e^{i\frac{2\rho}{\tau \hbar \sqrt{3}}\left[\arctan\left(\frac{2\tau x-1}{\sqrt{3}}\right)
		+\frac{\pi}{6}\right]}.\label{F1}
\end{eqnarray}
Moreover  from equation (\ref{moment}), we can deduce that
\begin{eqnarray}
	\frac{d}{d\rho} e^{-i\frac{2\rho}{\tau \hbar \sqrt{3}}\left[\arctan\left(\frac{2\tau x-1}{\sqrt{3}}\right)
		+\frac{\pi}{6}\right]}&=& -i\frac{2}{\tau \hbar \sqrt{3}}\left[\arctan\left(\frac{2\tau x-1}{\sqrt{3}}\right)
	+\frac{\pi}{6}\right] \cr&&\times e^{-i\frac{2\rho}{\tau \hbar \sqrt{3}}\left[\arctan\left(\frac{2\tau x-1}{\sqrt{3}}\right)
		+\frac{\pi}{6}\right]}.
\end{eqnarray}
This equation is equivalent to 
\begin{eqnarray}
	i\frac{\tau\hbar \sqrt{3}}{2}\frac{d}{d\rho}= \left[\arctan\left(\frac{2\tau x-1}{\sqrt{3}}\right)
	+\frac{\pi}{6}\right]=
	\left[\arctan\left(\frac{2\tau x-1}{\sqrt{3}}\right)
	+\arctan\left(\frac{1}{\sqrt{3}}\right)\right].\label{eq}
\end{eqnarray}
From the following relation \cite{19'}
\begin{eqnarray}
	\arctan\alpha +\arctan \beta=\arctan \left(\frac{\alpha+\beta}{1-\alpha\beta}\right),\quad \mbox{with}\quad \alpha\beta<1,
\end{eqnarray}
we deduce that
\begin{eqnarray}
	\tan \left[\arctan\left(\frac{2\tau x-1}{\sqrt{3}}\right)
	+\arctan\left(\frac{1}{\sqrt{3}}\right)\right]=\frac{\tau x \sqrt{3} }{2-\tau x}.
\end{eqnarray}
In equation (\ref{eq}), we can see that the position operator is represented as  \cite{11} 
\begin{eqnarray}
	\hat X&=&\frac{2}{\tau}\frac{\tan\left(i\frac{\tau\hbar \sqrt{3}}{2}\frac{d}{d\rho}\right)}{\sqrt{3}+\tan\left(i\frac{\tau\hbar \sqrt{3}}{2}\frac{d}{d\rho}\right)},\\
	\hat X\psi(\rho)&=&\frac{2}{\tau}\frac{\tan\left(i\frac{\tau\hbar \sqrt{3}}{2}\frac{d}{d\rho}\right)}{\sqrt{3}+\tan\left(i\frac{\tau\hbar \sqrt{3}}{2}\frac{d}{d\rho}\right)}\psi(\rho).
\end{eqnarray}
From the action of $\hat P$ on the  quasi representation (\ref{F1}) and using  equation (\ref{3}), we have
\begin{eqnarray}
	\hat P\psi(\rho)=\rho  \psi(\rho).
\end{eqnarray}
Note that in the limit $\tau\rightarrow 0$, we
 recover the corresponding ordinary quantum mechanics
 results in momentum space   (\ref{momentum}) 
 \begin{eqnarray}
 	\lim_{\tau\rightarrow 0}\hat X\psi(\rho)=i\hbar\frac{d}{d\rho}\psi(\rho)\quad \mbox{and}\quad 
 	\lim_{\tau\rightarrow 0}\hat P\psi(\rho)= \rho\psi(\rho).
 \end{eqnarray}

 
\section{Path integral  and propagator in position-deformed algebra}
From the path integrals within this position-deformed Heisenberg algebra, we construct the propagator depending on the position-representation and on the Fourier transform and its inverse representations.  We compute propagators and deduce the actions of a free particle and a harmonic oscillator as applications. 
\subsection{Path integral and propagator in position-space representation}
The Hamiltonian operator for a particle with mass $m$ living in one spatial
dimension is given by
\begin{eqnarray}
	\hat H=\frac{\hat P^2}{2m} +V(\hat X),
\end{eqnarray}
where V is the potential energy of the system. The time-dependent deformed Schrödinger equation in the position representation  is given by
\begin{eqnarray}\label{H1}
	\hat H  |\phi_\rho(t)\rangle =-\frac{\hbar^2}{2m} D_x^2 |\phi_\rho(t)\rangle +V(x)|\phi_\rho(t)\rangle= i\hbar \partial_t |\phi_\rho(t)\rangle.
\end{eqnarray}
The time-evolution process is described by
\begin{eqnarray}\label{Pro1}
	|\phi_\rho(t)\rangle= e^{-\frac{i}{\hbar}\hat H(t-t')} |\phi_\rho(t')\rangle,
\end{eqnarray}
Multiplication of $\langle x|$ from the left of the equation (\ref{Pro1}) gives
\begin{eqnarray}\label{F31}
	\phi_\rho (x,t)= \int_{-\infty}^{+\infty}\frac{dx'}{1-\tau x'+\tau^2 x'^2}K(x,t,x',t') \phi_\rho(x',t')
\end{eqnarray}
where $K$ is the kernel in Hamiltonian  or 
the amplitude for a particle to propagate from the state with position $x'$ to the state with position $x\,  (x>x')$ in a time interval  $\Delta t=t-t'$ \cite{20}  and it  is  defined as
\begin{eqnarray}\label{path1}
	K(x,t,x',t')=\langle x|e^{-\frac{i}{\hbar}\hat H(t-t')} |x'\rangle.
\end{eqnarray}
Splitting the interval $t-t'$ into N intervals of length $\epsilon=(t_k-t_{k-1})/N$ and inserting the completeness relations in (\ref{id2}) and (\ref{id}), the propagator (\ref{prop}) becomes
\begin{eqnarray}\label{prop}
	K(x,t,x',t')&=&\prod_{k=1}^{N-1} \left(\int_{-\infty}^{+\infty}\frac{dx_k}{1-\tau x_k+\tau^2 x_k^2}\right) \prod_{k=1}^{N} \left( \int_{-\infty}^{+\infty}\frac{d\rho_k}{\tau\hbar \sqrt{3}} \right)\cr&&\times \langle x_k|\rho_k\rangle\langle \rho_k|e^{-\frac{i}{\hbar}\epsilon\hat H}|x_{k-1}\rangle.
\end{eqnarray}
Recall that
\begin{eqnarray}
 \langle x_k|\rho_k\rangle&=& \phi_{\rho_k}(x_k) = \sqrt{\frac{\tau\sqrt{3}}{2\pi}} e^{\left(i\frac{2\rho_k}{\tau\hbar \sqrt{3}}\left[\arctan\left(\frac{2\tau x_k-1}{\sqrt{3}}\right)
 +\frac{\pi}{6}\right]\right)},\\
 \langle \rho_k|e^{-\frac{i}{\hbar}\epsilon\hat H}|x_{k-1}\rangle       &=& e^{-\frac{i}{\hbar}\epsilon H(\rho_k,x_{k-1})}  \langle \rho_k|x_{k-1}\rangle =e^{-\frac{i}{\hbar}\epsilon H(\rho_k,x_{k-1})} \phi_{\rho_k}^*(x_{k-1}).
\end{eqnarray}
Substituting these expressions into  equation (\ref{prop}) gives
\begin{eqnarray}
K(x,t,x',t')=\left[\prod_{k=1}^{N-1} \left(\int_{-\infty}^{+\infty}\frac{dx_k}{1-\tau x_k+\tau^2 x_k^2}\right)\right]\left[\prod_{k=1}^{N} \left(\frac{1}{\tau\hbar \sqrt{3}} \int_{-\infty}^{+\infty}d\rho_k \right)\right]
 e^{\frac{i}{\hbar}\epsilon\mathcal{S}_{disc}},
\end{eqnarray}
where the  discrete action $S_{disc}$ is
\begin{eqnarray}
S_{disc}&=& \sum_{k=1}^{N-1}\frac{2\rho_k}{\tau\sqrt{3}}\left[\frac{\arctan \left(\frac{2\tau x_{k}-1}{\sqrt{3}}\right) 
	-\arctan \left(\frac{2\tau x_{k-1}-1}{\sqrt{3}}\right)}{\epsilon}\right] -\sum_{k=1}^{N-1}H(\rho_k,x_{k-1})
\end{eqnarray}
Finally, we take the limit $N\rightarrow \infty$,  so that $\epsilon \rightarrow 0$. We obtain our final expression for the propagator as  follows

\begin{eqnarray}
	K(x,t,x',t')= \int_{-\infty}^{+\infty}\mathcal{D}x\mathcal{D}\rho e^{\frac{i}{\hbar} S },
\end{eqnarray}
where the  integration measures $\mathcal{D}x$ and $\mathcal{D}\rho$ are defined as
\begin{eqnarray}
	\mathcal{D}x=\lim_{N\rightarrow \infty}\prod_{k=1}^{N-1}\frac{dx_k}{1-\tau x_k+\tau^2 x_k^2}\quad \mbox{and}\quad \mathcal{D}\rho=\lim_{N\rightarrow \infty} \prod_{k=1}^{N}
	\left(\frac{d\rho_k}{\tau \hbar\sqrt{3}}\right).
\end{eqnarray}
and the continuous  action $S $ is given by
\begin{eqnarray}
 S \left[x(t),x(t')\right]=\int_{t'}^t d\nu  \left[\frac{\dot{x}(\nu)}{1-\tau x(\nu)+\tau^2 x^2(\nu)}\rho(\nu)-H(\rho(\nu),x(\nu))\right],
\end{eqnarray}
where $\dot{x}(\nu)= d x/d\nu  $. The stationary path is obtained by using the variational principle
\begin{eqnarray}\label{variational}
	\delta  S
	=\delta \int_{t'}^t d\nu L \left[\dot{x}(\nu),x(\nu)\right]=\int_{t'}^t d\nu \left(\frac{\partial L}{\partial x(\nu)}\delta x(\nu) +  \frac{\partial L}{\partial \dot{x}(\nu)}\delta \dot{x}(\nu)\right)=0,
\end{eqnarray}
where the Lagrangian $L$  of the system is given by
\begin{eqnarray}
	L \left[  \dot{x}(\nu),x(\nu)\right]= \frac{\dot{x}(\nu)}{1-\tau x(\nu)+\tau^2 x^2(\nu)}\rho(\nu)-H(\rho(\nu),x(\nu)).
\end{eqnarray}
The solutions of  equation (\ref{variational}) generates the  following differential equations
\begin{eqnarray}
\dot{x}=(1-\tau x+\tau^2 x^2)\frac{\partial H}{\partial \rho},\quad \dot{\rho}=-(1-\tau x+\tau^2 x^2)\frac{\partial H}{\partial x}.
\end{eqnarray}
 By taking the limit $\tau\rightarrow 0$, we recover the ordinary  Hamilton’s equations of motion.

%

\subsection{Path integral  and propagator in Fourier  transform and its inverse representions }
Using the generalized Fourier transform and its inverse representations (\ref{moment}), (\ref{Finverse}) and taking into account equation (\ref{F31}), we have
\begin{eqnarray}
	\psi (\rho,t)
	&=&\sqrt{\frac{\tau\sqrt{3}}{2\pi}} \int_{-\infty}^{+\infty}\frac{dx}{1-\tau  x+\tau^2 x^2} 
	e^{-i\frac{2\rho}{\tau \hbar \sqrt{3}}\left[\arctan\left(\frac{2\tau x-1}{\sqrt{3}}\right)
		+\frac{\pi}{6}\right]}\int_{-\infty}^{+\infty}\frac{K(x,t,x',t')}{1-\tau  x'+\tau^2 x'^2}dx'\cr&&
	\times \frac{1}{\hbar\sqrt{2\pi\tau\sqrt{3}}} \int_{-\infty}^{+\infty}d\rho'  e^{i\frac{2\rho'}{\tau \hbar \sqrt{3}}\left[\arctan\left(\frac{2\tau x'-1}{\sqrt{3}}\right)
		+\frac{\pi}{6}\right]}\psi(\rho',t').
\end{eqnarray}
This path integral can be  rewritten as follows
\begin{eqnarray}
		\psi (\rho,t)=\int_{-\infty}^{+\infty}d\rho'\mathcal{K}(\rho,t,\rho',t') 	\psi (\rho',t'),
\end{eqnarray}
where $ \mathcal{K}$ is the propagator in Fourier  transform and its inverse representions for a particle to go from a state $\psi (\rho') $
 to a state $\psi (\rho)$ in a time
interval $t-t'$ is
\begin{eqnarray}\label{key1}
	\mathcal{K}(\rho,t,\rho',t')&=&\frac{1}{2\hbar\pi}\int_{-\infty}^{+\infty}\frac{dx}{1-\tau x+\tau^2 x^2}\frac{dx'}{1-\tau x'+\tau^2 x'^2}\cr&&\times e^{-i\frac{2}{\tau \hbar \sqrt{3}}\left[\rho\arctan\left(\frac{2\tau x-1}{\sqrt{3}}\right)
			-\rho'\arctan\left(\frac{2\tau x'-1}{\sqrt{3}}\right)\right]}
		K(x,t,x',t'),\cr
		&=&\frac{1}{2\hbar\pi}\int_{-\infty}^{+\infty}\mathcal{D}x\mathcal{D}\rho\frac{dx}{1-\tau x+\tau^2 x^2}\frac{dx'}{1-\tau x'+\tau^2 x'^2}e^{\frac{i}{\hbar}\mathcal{S}},
\end{eqnarray}
 with $\mathcal{S}$ the   action  given by
 \begin{eqnarray}
 \mathcal{S}(\rho,t,\rho',t')= S-\frac{2}{\tau  \sqrt{3}}\left[\rho\arctan\left(\frac{2\tau x-1}{\sqrt{3}}\right)
 -\rho'\arctan\left(\frac{2\tau x'-1}{\sqrt{3}}\right)\right].	
 \end{eqnarray}

\subsection{Propagators for a free particle and for a harmonic oscillator }
In this  section, we  compute the propagator
in position-space (\ref{path1}) and the one in Fourier transform and its inverse representations  (\ref{key1}) for the Hamiltonians of a free particle  and a simple harmonic oscillator.  From these propagators, we deduce the actions of both systems.
\subsubsection{ A free particle}
\begin{figure}[htbp]
	\resizebox{0.8\textwidth}{!}{
		\includegraphics{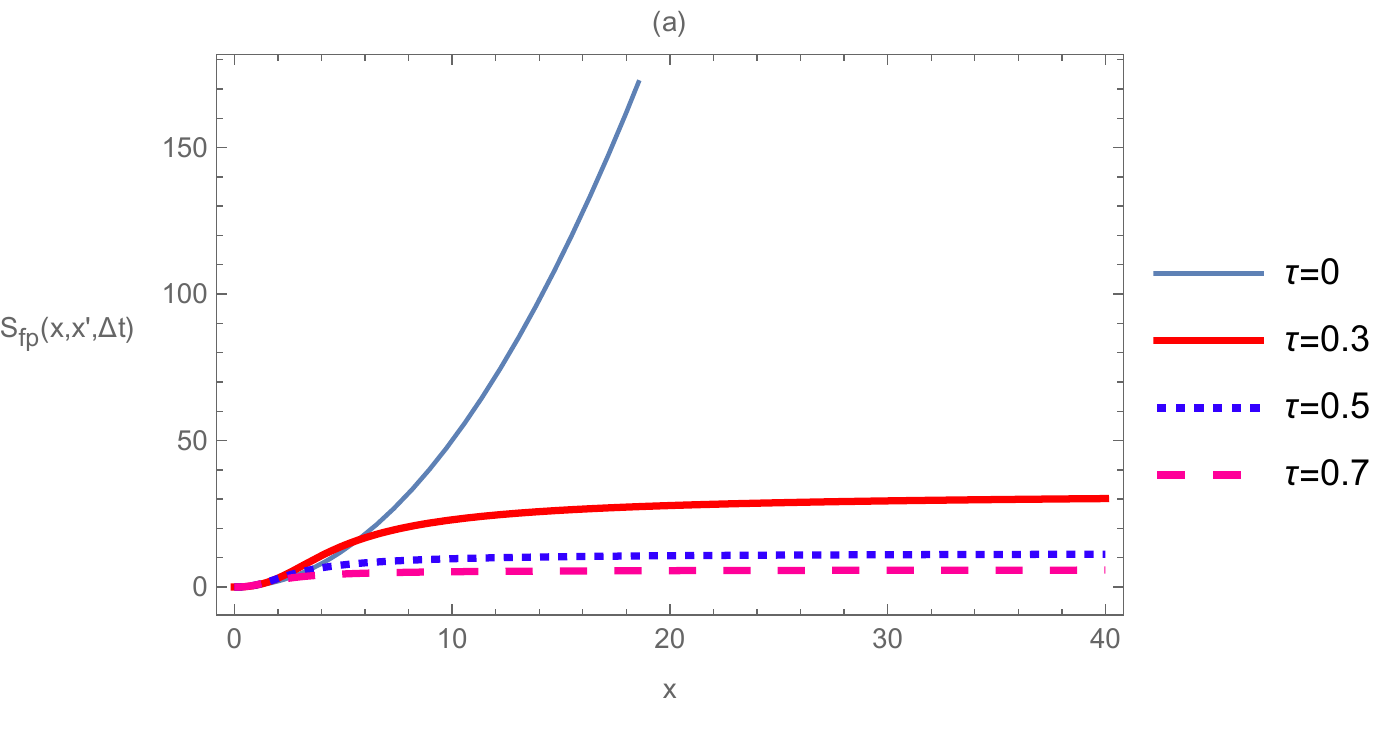}
	}
\resizebox{0.8\textwidth}{!}{
	\includegraphics{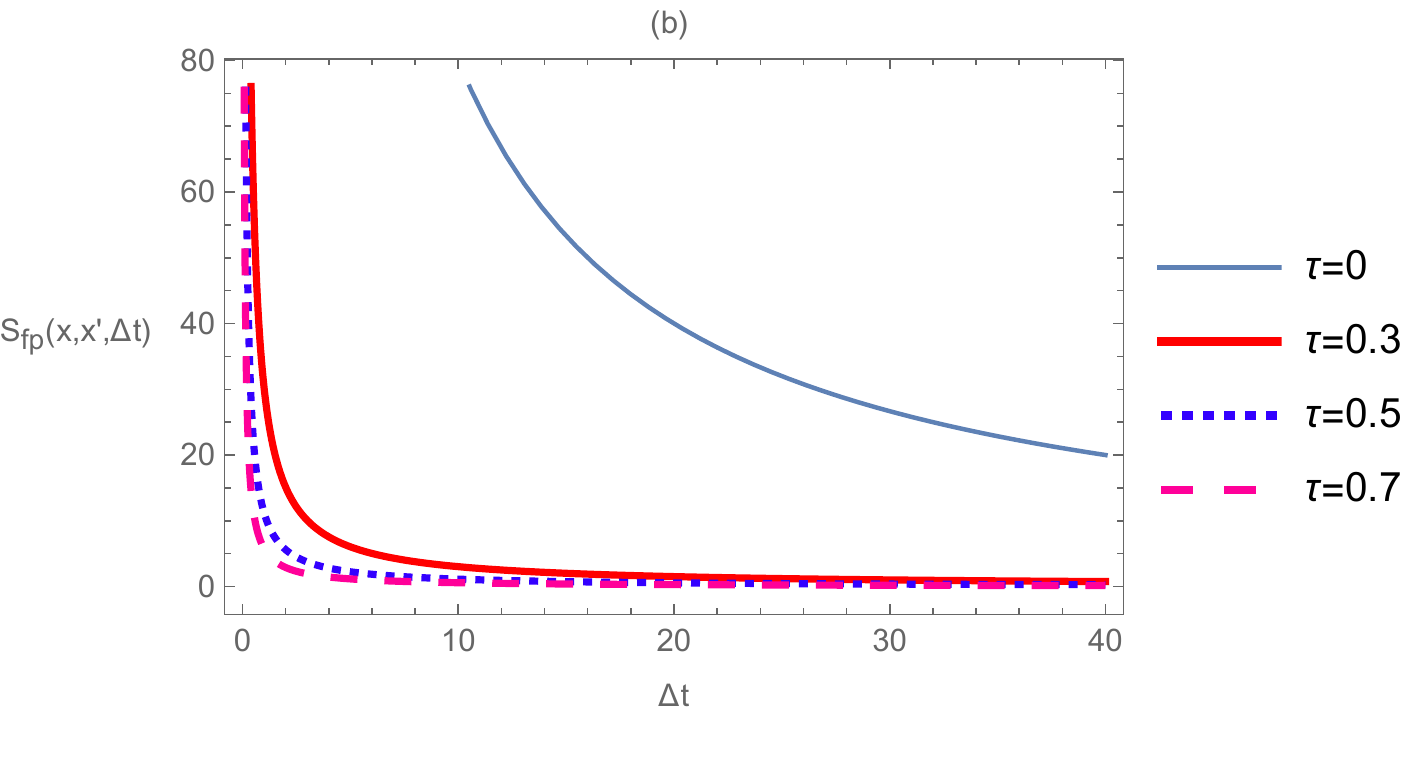}
}
\resizebox{0.8\textwidth}{!}{
	\includegraphics{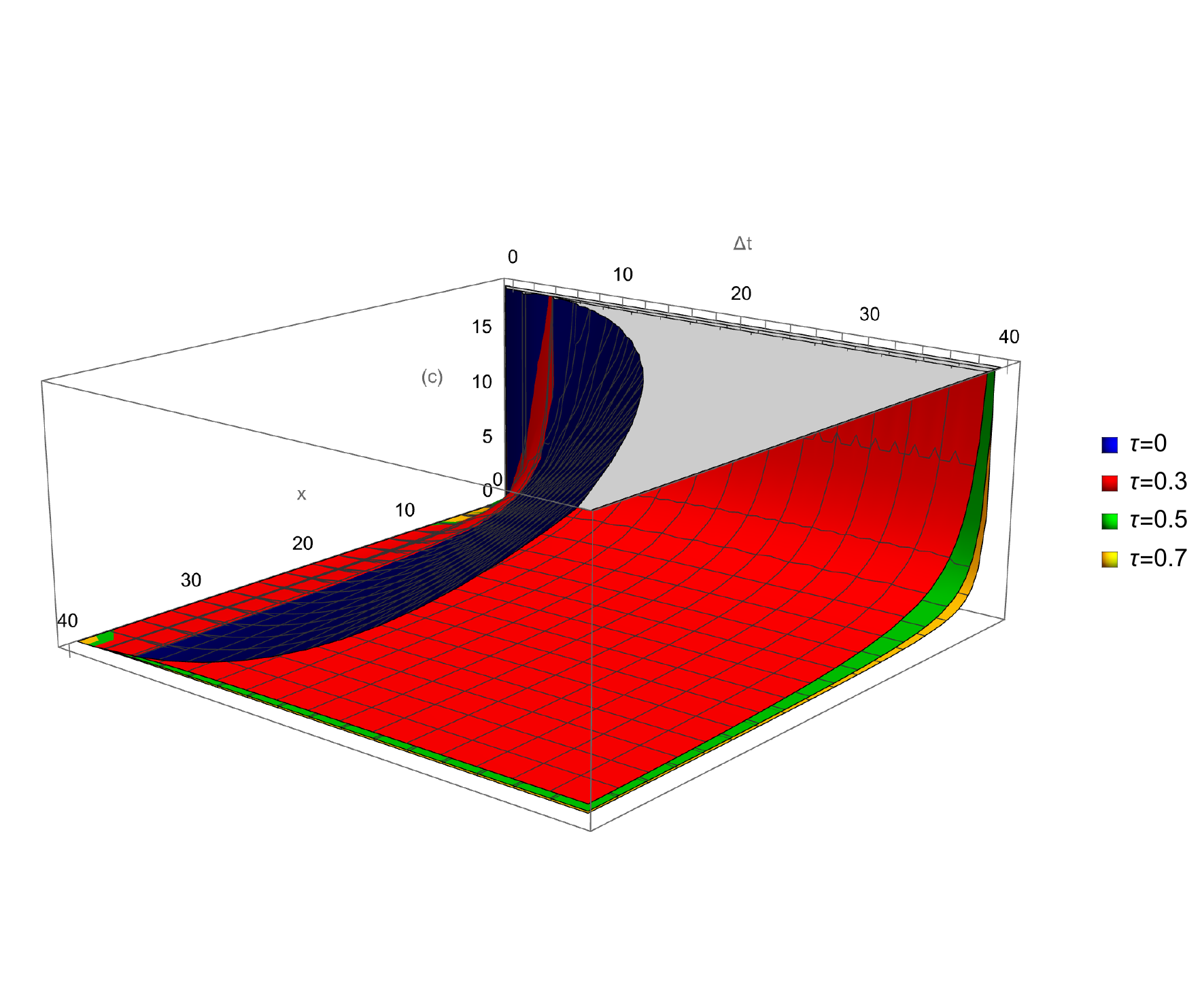}
}
		\caption{\it \small 
		Classical action of a free particle (\ref{daction}) versus the position $x$ and  the time $\Delta t$ for different values  of $\tau$ with $m =1$.
	}
	\label{fig1}       
\end{figure}

The free particle problem is defined by the Hamiltonian given by
\begin{eqnarray}
	\hat H_{fp}=\frac{\hat P^2}{2m}.
\end{eqnarray}
The propagator in position-represention in the  time interval $\Delta t =t-t'$ is  given by
\begin{eqnarray}
		K_{fp}(x,x',\Delta t) &=&\langle x|e^{-\frac{i}{\hbar}\frac{\hat P^2}{2m}\Delta t}|x'\rangle\cr
	 &=&\frac{1}{\tau\hbar \sqrt{3}}\langle x|\int_{-\infty}^{+\infty}d\rho e^{-\frac{i}{\hbar}\frac{\hat P^2}{2m}\Delta t}|\rho\rangle\langle \rho|x'\rangle\cr
	 &=&\frac{1}{\tau\hbar \sqrt{3}}\int_{-\infty}^{+\infty}d\rho e^{-\frac{i}{\hbar}\frac{\rho^2}{2m}\Delta t} \langle x|\rho\rangle\langle \rho|x'\rangle\cr
	  &=&\int_{-\infty}^{+\infty}\frac{d\rho}{2\pi\hbar} e^{\left(i\frac{2\rho}{\tau\hbar \sqrt{3}}\left[\arctan\left(\frac{2\tau x-1}{\sqrt{3}}\right)
	  	-\arctan\left(\frac{2\tau x'-1}{\sqrt{3}}\right)\right]-\frac{i}{\hbar}\frac{\rho^2}{2m}\Delta t\right)}.\label{y1}
\end{eqnarray}
Completing this Gaussian integral (\ref{y1}), we have
\begin{eqnarray}
K_{fp}(x,x',\Delta t) =	\sqrt{\frac{ m}{2\pi\hbar i\Delta t}} 	e^{i\frac{2m}{\hbar3\tau^2 \Delta t}\left[\arctan\left(\frac{2\tau x-1}{\sqrt{3}}\right)
	-\arctan\left(\frac{2\tau x'-1}{\sqrt{3}}\right)\right]^2}.
\end{eqnarray}
Thus the deformed classical action is given by
\begin{eqnarray}\label{daction}
S_{fp} =\frac{2m}{3\tau^2 \Delta t}\left[\arctan\left(\frac{2\tau x-1}{\sqrt{3}}\right)
-\arctan\left(\frac{2\tau x'-1}{\sqrt{3}}\right)\right]^2.
\end{eqnarray}
The limit $\tau\rightarrow 0$,  the latter propagator  properly
reduces to the well-known  result in ordinary quantum mechanics for a free particle \cite{21,22}  that is
\begin{eqnarray}
	\lim_{\tau\rightarrow 0}	K_{fp}(x,x',\Delta t)&=&K_{fp}^0(x,x',\Delta t)=\sqrt{\frac{ m}{2\pi\hbar i\Delta t}} e^{\frac{i}{\hbar}\frac{m(x-x')^2}{2\Delta t}},
\end{eqnarray}
 and the  corresponding classical  action  is given by
 \begin{eqnarray}
 	\lim_{\tau\rightarrow 0} S_{fp}= S_{fp}^0= \frac{m}{2}\frac{(x-x')^2}{\Delta t}=KE,
 \end{eqnarray}
where KE is the kenetic energy of the particle.
Also, it is straightforward to show  the following relations
\begin{eqnarray}
	K_{fp}(x,x',\Delta t)\leq K_{fp}^0(x,x',\Delta t)\implies S_{fp}\leq S_{fp}^0, 
\end{eqnarray}
which indicate that the propagator and  the actions of  free particles  are dominated by standard ones free of gravity deformation.\\
Figure (\ref{fig1}) illustrates the deformed action (\ref{daction}) of the free particle  versus the position  $x$ (with $x'=0$) and the time $\Delta t$ for fixed values of parameter $\tau$. Figure $1a$ shows that, for any  $\Delta t>0$  the values of the deformed  action $ S_{fp}$ over the position decrease from  the non deformed action $ S_{fp}^0$ as one increases the quantum gravitational parameter $\tau$. As it can also be seen in Figure $1b$, $ S_{fp}$ decreases over the time $\Delta t$ for large distance ($\Delta x=40 $). $S_{fp}$ rapidily decreases  as one increases the parameter of $\tau$. Figure $1c$ clearly shows this for the simultaneous variation of $S_{fp}$ in time $\Delta t$ and position $x$. These results indicate that quantum gravitational effects in this space  shorten the paths of particles, allowing them to move from one point to another in a short time. In one way or another,  because the classical action of free particle has the dimenson of  energy (KE). These results can be understood as free particles use low  energies to travel  quickly in this  deformed space.
 This strengthens the claim that the position deformed algebra (\ref{alg2})  induces strong quantum gravitational  fields with features close to the classical ones \cite{1}.

The propagator for the Fourier  transform and its inverse representions is given by
\begin{eqnarray}
		\mathcal{K}(\rho,\rho',\Delta t)&=& 
			\frac{1}{2\hbar\pi}\sqrt{\frac{ m}{2\hbar\pi i\Delta t}}\int_{-\infty}^{+\infty}\frac{dx}{1-\tau x+\tau^2 x^2}\int_{-\infty}^{+\infty}\frac{dx'}{1-\tau x'+\tau^2 x'^2}\cr&&\times e^{-i\frac{2}{\tau \hbar \sqrt{3}}\left[\rho\arctan\left(\frac{2\tau x-1}{\sqrt{3}}\right)
				-\rho'\arctan\left(\frac{2\tau x'-1}{\sqrt{3}}\right)\right]}\cr&&
			\times e^{i\frac{2m}{\hbar3\tau^2 \Delta t}\left[\arctan\left(\frac{2\tau x-1}{\sqrt{3}}\right)
					-\arctan\left(\frac{2\tau x'-1}{\sqrt{3}}\right)\right]^2}.
\end{eqnarray}
The corresponding action is given by
\begin{eqnarray}
	\mathcal{S}_{fp}= S_{fp}- \frac{2}{\tau  \sqrt{3}}\left[\rho\arctan\left(\frac{2\tau x-1}{\sqrt{3}}\right)
	-\rho'\arctan\left(\frac{2\tau x'-1}{\sqrt{3}}\right)\right].
\end{eqnarray}
%

\subsubsection{ A simple Harmonic  oscillator}

The simple harmonic oscillator problem is defined by the Hamiltonian
\begin{eqnarray}
	\hat H=\frac{\hat P^2}{2m}+\frac{1}{2}m\omega^2\hat X^2.
\end{eqnarray}
The propagator in position representation is given by 
\begin{eqnarray}
	K_{ho}(x,x',\Delta t) &=&\langle x|e^{-\frac{i}{\hbar}\left(\frac{\hat P^2}{2m}+\frac{1}{2}m\omega^2\hat X^2\right)\Delta t}|x'\rangle\cr
	&=&\frac{1}{\tau\hbar \sqrt{3}}\langle x|e^{-\frac{i}{2\hbar}m\omega^2 x'^2\Delta t}\int_{-\infty}^{+\infty}d\rho e^{-\frac{i}{\hbar}\frac{\hat P^2}{2m}\Delta t}|\rho\rangle\langle \rho|x'\rangle\cr
	&=&\int_{-\infty}^{+\infty}\frac{d\rho}{2\pi\hbar} \cr && \times e^{\left(i\frac{2\rho}{\tau\hbar \sqrt{3}}\left[\arctan\left(\frac{2\tau x-1}{\sqrt{3}}\right)
		-\arctan\left(\frac{2\tau x'-1}{\sqrt{3}}\right)\right]-\frac{i}{\hbar}
		\left(\frac{ \rho^2}{2m}+\frac{1}{2}m\omega^2 x'^2\right)\Delta t\right)}.\label{y3}
\end{eqnarray}
Computing the Gaussian integral (\ref{y3}), we have
\begin{eqnarray}
	K_{ho}(x,x',\Delta t) =	\sqrt{\frac{ m}{2\pi\hbar i\Delta t}} 	e^{i\frac{2m}{\hbar3\tau^2 \Delta t}\left[\arctan\left(\frac{2\tau x-1}{\sqrt{3}}\right)
		-\arctan\left(\frac{2\tau x'-1}{\sqrt{3}}\right)\right]^2-\frac{i}{2\hbar}m\omega^2 x'^2\Delta t},
\end{eqnarray}
  and the corresponding deformed classical action is given by
  \begin{eqnarray}
  	S_{ho} =\frac{2m}{3\tau^2 \Delta t}\left[\arctan\left(\frac{2\tau x-1}{\sqrt{3}}\right)
  	-\arctan\left(\frac{2\tau x'-1}{\sqrt{3}}\right)\right]^2-\frac{1}{2}m\omega^2 x'^2\Delta t.
  \end{eqnarray}
  At the limit $ \tau\rightarrow 0$, we recover  the ordinary propagator and the classical action of the simple harmonic oscillator \cite{21,22}
\begin{eqnarray}
	\lim_{\tau\rightarrow 0}	K_{ho}(x,x',\Delta t)&=& 	K_{ho}^0(x,x',\Delta t)=\sqrt{\frac{ m}{2\pi\hbar i\Delta t}} e^{\frac{i}{\hbar}\left(\frac{m(x-x')^2}{2\Delta t}-\frac{1}{2}m\omega^2 x'^2\Delta t\right)},\cr
		\lim_{\tau\rightarrow 0} S_{ho}&=& S_{ho}^0= \frac{m(x-x')^2}{2\Delta t}-\frac{1}{2}m\omega^2 x'^2\Delta t=E_m,
\end{eqnarray} 
where $E_m$ is the mechanical energy of a simple  harmonic mechanics. Like in the prior instance (\ref{action1}), It is simple to demonstrate that 
\begin{eqnarray}\label{action1}
	K_{ho}(x,x',\Delta t)\leq K_{ho}^0(x,x',\Delta t)\implies S_{ho}\leq S_{ho}^0. 
\end{eqnarray}
In more general case, we can see that the harmonic oscillator potential  does not affect the motion of the deformed motion of the free particle such as 
\begin{eqnarray}
	K_{ho}\approx 	K_{fp}\leq K_{fp}^0 \approx	K_{ho}^0 \implies S_{ho}\approx	S_{fp}\leq S_{fp}^0 \approx	S_{ho}^0. 
\end{eqnarray}

The propagator in Fourier  transform and its inverse representions is given by 
\begin{eqnarray}
	\mathcal{K}(\rho,\rho',\Delta t)&=& 
	\frac{1}{2\pi}\sqrt{\frac{ m}{2\pi\hbar i\Delta t}}\int_{-\infty}^{+\infty}\frac{dx}{1-\tau x+\tau^2 x^2}\int_{-\infty}^{+\infty}\frac{dx'}{1-\tau x'+\tau^2 x'^2}\cr&&\times e^{-i\frac{2}{\tau \hbar \sqrt{3}}\left[\rho\arctan\left(\frac{2\tau x-1}{\sqrt{3}}\right)
		-\rho'\arctan\left(\frac{2\tau x'-1}{\sqrt{3}}\right)\right]}\cr&&
	\times e^{i\frac{2m}{\hbar3\tau^2 \Delta t}\left[\arctan\left(\frac{2\tau x-1}{\sqrt{3}}\right)
		-\arctan\left(\frac{2\tau x'-1}{\sqrt{3}}\right)\right]^2-\frac{i}{2\hbar}m\omega^2 x'^2\Delta t},
\end{eqnarray}
 and its action is  given by
 \begin{eqnarray}
 	\mathcal{S}_{ho}&=& S_{ho}   - \frac{2}{\tau  \sqrt{3}}\left[\rho\arctan\left(\frac{2\tau x-1}{\sqrt{3}}\right)
 	-\rho'\arctan\left(\frac{2\tau x'-1}{\sqrt{3}}\right)\right].
  \end{eqnarray}

\section{Conclusion}
We have constructed path integrals in Euclidean position representation and in Fourier transform and its inverse representations within a position-deformed Heisenberg algebra. We have derived from these path integrals the propagators and the corresponding classical  actions. The classical equations of motion are obtained by the principle of least action. The Hamiltonians of a free particle and a simple harmonic oscillator are used as examples to compute the propagators and the actions in position representation and in Fourier transform and inverse representations. We have  shown through these that the propagators and the actions of these systems in position space representation are properly bounded by the well-known results in the $\tau\rightarrow 0$ limit. These mathematical results have been confirmed by the
numerical investigations of the classical action of these  systems. We have observed that simultaneous variation of the action  in time and in position rapidily decreases as  one increases the parameter of quantum gravity $\tau$. This suggests that quantum gravity in this space bends particle pathways, allowing them to travel fast from one point to the next. The propagators  for Fourier transform and its inverse representations for both  systems are given as integral expressions and  we have deduced the corresponding actions.

 In this work, we have constructed  path integrals in the deformed Heisenberg algebra from the Schrödinger equation. One  can  extend this work on the stochastic path integrals using the Fokker-Planck equation \cite{23,24,25,26} or to derive the Black–Scholes pricing kernel  from  the Black–Scholes equation \cite{27}.

\section*{Acknowledgments}
LML acknowledges support from DAAD (German Academic Exchange Service) under the DAAD postodoctoral in region grant



\begin{thebibliography}{99}
	
\bibitem{1} L. Lawson, {Position-dependent mass in strong quantum gravitational background fields}, J. Phys. A: Math. Theor. {\bfseries 55}, 105303 (2022)


\bibitem{3}    A. Kempf, G. Mangano and R. Mann,  Hilbert space representation
	of the minitial length uncertainty relation,  Phys. Rev. D {\bfseries 52},  1108  (1995)

\bibitem {4} A. Kempf, G. Mangano,  Minimal length uncertainty relation and ultraviolet regularization, Phys. Rev. D. {\bf 55}  7909-7920 (1997). 

\bibitem{5}  Y. Sabri and K. Nouicer, Phase transitions of a GUP-corrected Schwarzschild black hole within isothermal cavities, Class. Quant. Grav. {\bf 29} , 215015 (2012)


\bibitem{6} A. Ali, S. Das and E. Vagenas, Discreteness of space from the generalized uncertainty principle, Phys. Lett.B  {\bf 678}, 497 (2009)

\bibitem{7} S. Das, E. Vagenas and A. Ali, Discreteness of space from GUP II: Relativistic wave equations, Phys. Lett. B, {\bf 690}, 407 (2010)

\bibitem{8} Pouria Pedram, A higher order GUP with minimal length uncertainty and maximal momentum, Physics Letters B {\bf 714},  317-323  (2012)

\bibitem{9} Pouria Pedram,  A higher order GUP with minimal length uncertainty and maximal momentum II, Physics Letters B {\bf 718},  638–645  (2012)


\bibitem{10} F. Scardiglia and R. Casadio,  Gravitational tests of the Generalized Uncertainty Principle, Eur. Phys. J. C  {\bfseries 75}, 425 (2015).

\bibitem{11} K. Nozari and A. Etemadi, Minimal length, maximal momentum and Hilbert space representation of quantum mechanics, Phys. Rev. D {\bf 85}, 104029    (2012)


\bibitem{12} A. Tawfik and A. Diab,  A review of the generalized uncertainty principle, Rep. Prog. Phys. {\bf 78}, 126001 (2015)



\bibitem{13}   L. Lawson, L. Gouba and G. Avossevou, 	Two-dimensional noncommutative gravitational quantum well, J. Phys A: Math. Theor {\bfseries 50},   475202  (2017)

\bibitem{14} W. Sang Chung and H. Hassanabadi, A new higher order GUP: one dimensional quantum system, Eur. Phys. J.C {\bfseries 79}, 213 (2019)

\bibitem{15} L. Lawson,  Minimal and maximal lengths from position-dependent
	noncommutativity, J. Phys. A: Math. Theor. {\bf 53}, 115303 (2020) 
	
	
	\bibitem{16} R. Bernardo and J. Esguerra, Euclidean path integral formalism in deformed space with minimum
	measurable length, J. Math. Phys. {\bfseries 58}, 042103 (2017)
	

\bibitem{17} S. Bhattacharyya and S. Gangopadhyay, Path-integral action in the generalized uncertainty principle framework, Phys. Rev. D {\bfseries 104}, 026003  (2021) 


\bibitem{18} K. Nouicer, Coulomb potential in one dimension with minimal length: A path integral approach,  J. Math. Phys. {\bfseries 48}, 112104 (2007)


\bibitem{19} P. Valtancoli, Path integral and noncommutative Poisson brackets, J. Math. Phys. {\bfseries 56}, 063501 (2015)

\bibitem{19'}  I. Gradshteyn and M. Ryzhik, Table of Integrals, Series, and Products, 8th ed. (Academic Press, San Diego, California,
USA, 2015).


\bibitem{20} S. Das and  S. Pramanik, Path integral for nonrelativistic generalized uncertainty principle corrected Hamiltonian, Phys. Rev. D {\bfseries 86}, 085004 (2012)

\bibitem{21} H. Klein, Path Integral in Quantum Statistics and Polymer Physics, World Scientific Singapore, 1990

\bibitem{22}
  D. C. Khandekar, S. V. Lawande, and K. V. Bhagwat, Path Integrals Methods and 
 Their Application, World Scientific Singapore,1993
 
 \bibitem{23} P. Bressloff, Coherent spin states and stochastic hybrid path integrals, J. Stat. Mech. (2021) 043207
 
 \bibitem{24} P. Bressloff, Construction of stochastic hybrid path integrals using operator methods,  J. Phys. A: Math. Theor. {\bfseries 54}, 185001 (2021)
 
 \bibitem{25} J. Vastola and W. Holmes, Stochastic path integrals can be derived like quantum mechanical
 path integrals, arXiv:1909.12990 [cond-mat.stat-mech]
 
 \bibitem{26} B. da Costa, I. Gomez, and E. Borges, Deformed Fokker-Planck equation: Inhomogeneous medium with a position-dependent mass, Phys. Rev. E {\bfseries 102}, 062105 (2020) 
 \bibitem{27} B. Baaquie, Quantum Finance, Path Integral and Hamiltonians for  options and interest rates, Cambridge University Press
 The Edinburgh Building, Cambridge CB2 8RU, UK
 
       













\end{thebibliography}
\end{document}